\documentclass[twocolumn,showpacs,preprintnumbers,amsmath,amssymb]{revtex4}

\usepackage{graphicx}

\usepackage{dcolumn}

\usepackage{bm}

\usepackage{color}
\begin{document}

\preprint{APS/123-QED}

\title{Tipping without Flipping: A Novel Metastable ``Tilted" State in Anisotropic Ferromagnets in External Fields}
\author{Leiming Chen}
\address{College of Science, China University of Mining and Technology, Xuzhou 221116, P.R. China}
\author{John Toner}
\address{Department of Physics and Institute of Theoretical Science, University of Oregon, Eugene, OR 97403} \date{\today}

\begin{abstract}
We show that in suitable anisotropic ferromagnets, both stable and metastable ``tilted'' phases occur, in which the magnetization ${\vec M}$ makes an angle between zero and $180$ degrees with the externally applied ${\vec H}$. Tuning either the magnitude of the external field or the temperature can lead to continuous transitions between these states. A unique feature is that one of these transitions is between two {\it metastable} states.
Near the transitions the longitudinal
susceptibility becomes anomalous with
an exponent which has an {\it exact} scaling relation with the critical exponents.
\end{abstract}
\pacs{75.30.Kz, 75.30.Gw, 75.50.Gg, 75.10.-b} 
\maketitle 
Anisotropic anti-ferromagnets\cite{MEFisher76, Aharony} have long been known to have very rich phase diagrams, including  phases exhibiting Ising, $XY$, and both types of order simultaneously.

Less attention seems to have been focussed on the ferromagnetic case. Here we analyze that case, and show that it, too, can exhibit phases with simultaneous Ising and $XY$ order. In this case, such ``mixed" order implies extremely novel ``tilted" phases,  in which the magnetization ${\vec M}$ makes an angle between zero and $180$ degrees with the externally applied ${\vec H}$. These phases can be both stable and metastable.

\begin{figure}
\includegraphics[width=0.30\textwidth,angle=-90]{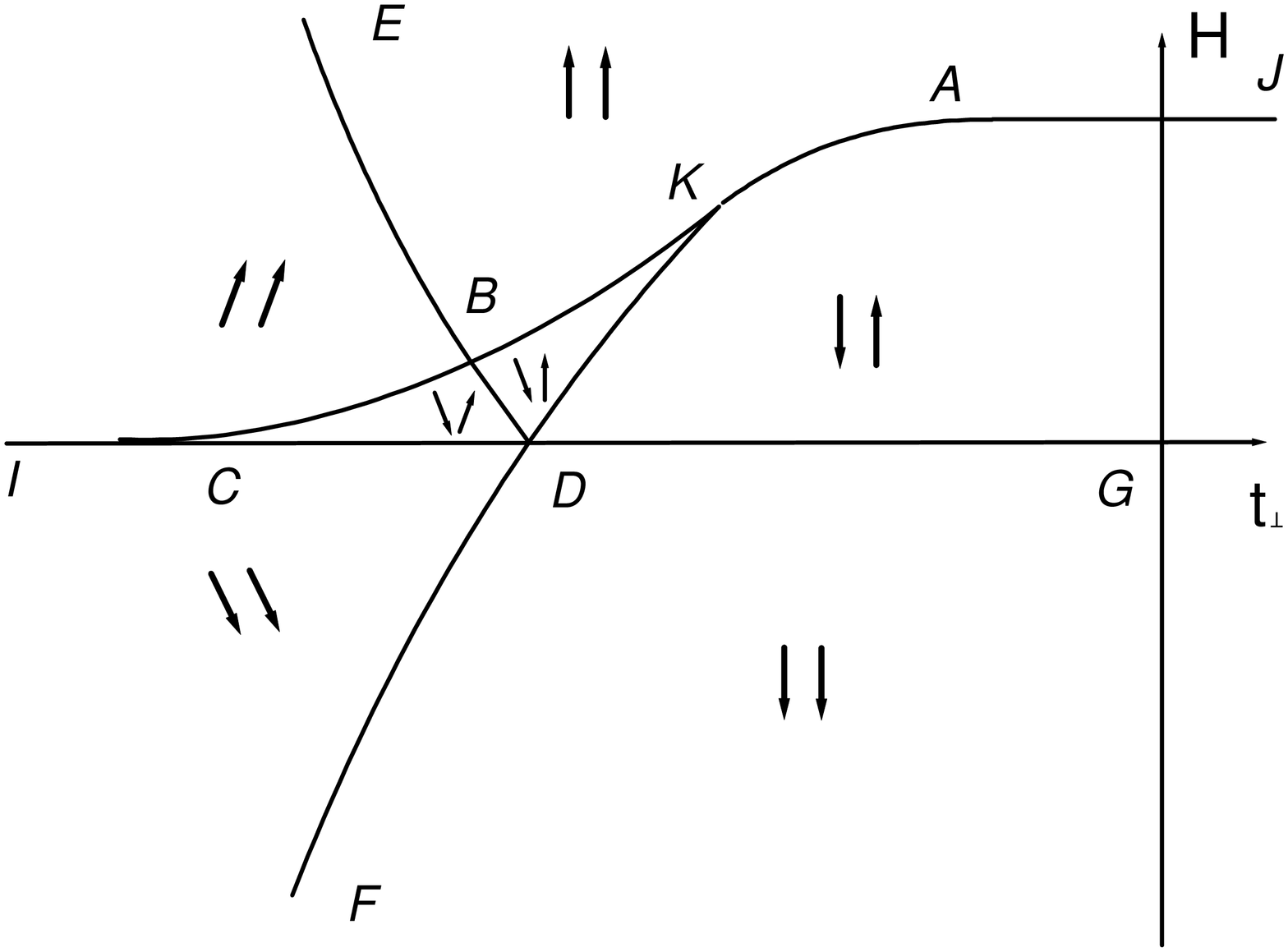}
\caption{\label{fig: Figure1}Phase diagram for hexagonal ferromagnets below the Curie temperature in the presence of an external field ${\vec H}$ perpendicular to the hexagonal planes (this direction will hereafter be called the ``${\hat z}$-axis''). Each phase is identified by two arrows. The right arrow denotes the orientation of the magnetization in the true equilibrium state, and the left one denotes the orientation in the state actually reached upon increasing $H_z$ from large negative values; when this differs from the true equilibrium state, the state is metastable.
The locus $KD$ is a ``meta-critical'' line separating two metastable states, while the locii $BE$ and $DF$ are ``true'' critical lines separating distinct equilibrium states.} \end{figure}

Our results are summarized in Fig. \ref{fig: Figure1}, which illustrates both the equilibrium and metastable phases of a suitable hexagonal ferromagnetic crystal in the $H$-$t_{\perp}$ plane, where $H$ is the external magnetic field, and $t_{\perp}$ is a phenomenological parameter that increases monotonically with temperature $T$.
``Suitable'', in this context, means the crystal field obeys certain conditions that we will specify more precisely later. For now, we simply point out that these conditions prove to be ``generic'': that is, they do {\it not} require ``fine-tuning'' of any material parameters.
This
does not mean that {\it all} hexagonal crystals will exhibit the phase diagram Fig. \ref{fig: Figure1}; it simply means that {\it some} of them should.

Metastable phases,  of course, depend on the sample history. In Fig. \ref{fig: Figure1}, we have assumed that the system starts in an ordered state with both the external field ${\vec H}$ and the magnetization ${\vec M}$ large, and in the $-{\hat z}$ direction, where ${\hat z}$ is the unit vector normal to the hexagonal planes.
Keeping temperature fixed,
and
the external field ${\vec H}$ along the ${\hat z}$ axis, the component $H_z$ of ${\vec H}$ along ${\hat z}$ is then varied from large negative to large positive values.

In the region below the locus $FDG$ and above the locus $EBKAJ$, there is no metastability, and the equilibrium state is just the conventional one, with the magnetization ${\vec M}\parallel {\vec H}$, the external field.
The region $JAKDG$ is likewise quite familiar: here, the equilibrium state remains ${\vec M}\parallel {\vec H}$, while the metastable state is the conventional one with ${\vec M}\parallel - {\vec H}$.

All of the other regions of the phase diagram exhibit tilted phases either in the equilibrium or the metastable state, or both.
In the region $BKD$, the tilted phase is metastable; the equilibrium state still has ${\vec M}\parallel {\vec H}$.
In the regions $CDF$ and $CBE$, there is no metastability, the equilibrium states are tilted; while in $CBD$, both the metastable and equilibrium states are tilted, but the components of ${\vec M}$ along the external field are of {\it opposite} sign!

The locii $DF$ and
$EB$ are equilibrium tilting transitions, while the locus $KD$ is a very strange one indeed: it represents a purely {\it metastable} transition: i.e., a continuous transition between two metastable states.

For the sample history
we specified, the locus $BD$ is not a transition line if the system is trapped in the metastable states. However, if the system is allowed to equilibrate, it is also a equilibrium transition.

In all of the tilted states, the projection of the tilted magnetization onto the hexagonal planes always lies along one of six six-fold symmetry related directions.

For a part of the parameter space of our model, the tilting is continuous, and belongs to the universality class of the three-dimensional $XY$ model\cite{lubensky_book}; that is, for $H\to H_c(t_\perp)$, where $H_c(t_\perp)$ is the $t_\perp$-dependent critical field $H$ at which the tilting transition occurs (i.e., value of $H$ on one of the tilting locii $DF$, $EB$, and $KD$ just discussed), the tilt angle $\theta$ is given by \begin{eqnarray} \theta\sim |H-H_c|^{\beta}\, , \label{tilt_angle} \end{eqnarray} where the universal exponent $\beta = 0.3485 \pm 0.0002$\cite{Camp} is  the order parameter exponent for the three-dimensional $XY$ model.

The tilting transition can also be crossed by varying temperature. In this case, $H$ and $H_c$ are replaced in Eq. (\ref{tilt_angle}) with $T$ and $T_c$, respectively, with $T_c$ being the temperature on the tilting locii.
Note that, in contrast to the usual ferromagnet-paramagnet critical point, temperature $T$ and external field $H $ are equivalent here, in the sense just described.

Since the projection ${\vec M_\perp}$ of the magnetization ${\vec M}$ perpendicular to the applied field is the order parameter for this tilting transition, both the associated susceptibility (namely, the uniform transverse susceptibility) $\chi_{\perp}$ and the correlation length $\xi$ for correlations of $\vec{M}_\perp$ diverge near the tilting transition, according to the laws:
\begin{eqnarray}
\chi_{\perp}\sim |H-H_c|^{-\gamma},~~~
\xi\sim |H-H_c|^{-\nu},
\label{susc_corr_length}
\end{eqnarray}
where the critical exponents $\gamma=1.3177\pm0.0005$ and
$\nu=.67155\pm0.00027$ are
respectively the universal susceptibility and correlation length  exponents of the three-dimensional $XY$ model\cite{Camp}.

In addition, the generalized {\it longitudinal} susceptibility is renormalized by the critical fluctuations and becomes wavelength-dependent. It displays a weak anomaly \begin{eqnarray}
\chi_z(\vec{q})&\approx&\mbox{constant}-
\left\{
\begin{array}{ll}
C_{\pm}q^{-\alpha/\nu} , &q\gg\xi^{-1}\\ C_{\pm}|H-H_c|^{-\alpha} , &q\ll\xi^{-1} \end{array}\right.
\label{Weak}
\end{eqnarray}
where $\alpha=-0.0146\pm 0.0008$ is the universal specific heat exponent of the three-dimensional $XY$ model.
$C_{\pm}$ are non-universal positive constants for $H$ above and below $H_c$, respectively. Their ratio $C_+\over C_-$ is universal.

We have also investigated these phenomena for other crystal lattice symmetries. For a cubic ferromagnetic crystal we find that if the external field is along one of the cubic axes (e.g., (100)), a {\it continuous} tilting transition is possible. The universality class of the transition is again that of the three-dimensional $XY$ model, and the longitudinal susceptibility is again  given by (\ref{Weak}).

If the external field is along one of the body diagonal directions (e.g., (111)), the tilting transition is in the universality class of the three-dimensional three-state Potts model, which is believed \cite{FisherPotts} to be first order.

In the case of an orthorhombic crystal, when the external field is along one of the three non-equal primary axes, the tilting transition can be continuous. Furthermore, it can happen between
metastable as well as equilibrium states. The universality class of this transition is that of the three-dimensional Ising model. Near the critical point $\chi_z$ displays a divergence of the
form (\ref{Weak}), but with $C_{\pm} < 0$,
$\nu = 0.630\pm0.001$, and $\alpha=0.109\pm 0.004$\cite{Ising}, where $\nu$ and $\alpha$ are respectively the universal correlation length and  specific heat exponents of the three-dimensional Ising model.

Our model for an anisotropic hexagonal ferromagnet in an external field $\vec{H}$
is:
\begin{eqnarray}
F &=& {1\over 2}\int d^dr~\left[g_z M_z^2 + t_{\perp}|\vec{M}_{\perp}|^2 +
C|\vec{\nabla}\vec{M}|^2+u_z M_z^4\right.\nonumber\\
&~&\left.+u_{\perp} |\vec{M}_{\perp}|^4
+u_{\perp z}|\vec{M}_{\perp}|^2M_z^2-2HM_z\right],
\label{H with field}
\end{eqnarray}
where $\perp$ denotes the $x-y$ plane, taken to be the plane of hexagonal symmetry, and the positive direction of the field is along $\hat{z}$. This model with $H = 0$ was first studied by Fisher et.al.\cite{MEFisher76} as a model for {\it antiferromagnets} in uniaxial crystals. It includes all terms to fourth order in $\vec M$ allowed by the hexagonal symmetry.
There are  {\it sixth} order in $\vec M$ terms  allowed by the hexagonal symmetry that break the continuous rotation invariance of this model in the $\perp$-plane down to six-fold rotational invariance.
Such terms pick out six equivalent preferred directions within the $\perp$-plane for the tilting, but do not affect either the topology of the phase diagram in the figure, or the universality classes of any of the various transitions therein.

In what follows we will consider only the case $g_z<0$.
Moreover, for simplicity we restrict ourselves to
the region of parameter space $u_{\perp,z}>0$, $0<u_{\perp z}<2\sqrt{u_{\perp} u_z}$.

We take the initial field to be so strong (i.e., a large negative $H$) that in the ground state the nonzero magnetization points along $-\hat{z}$.

We will begin by treating this model in Landau theory, in which we find the state of the system by minimizing this Landau free energy Eq. (\ref{H with field}). Expanding it around the ground state by writing $\vec{M} = (-M_0 + \delta M_z)\hat{z} + \vec{M}_{\perp}$ where $M_0$ satisfies $M_0g_z+2u_zM_0^3+H=0$, we obtain \begin{eqnarray} F &=& {1\over 2}\int d^dr~\left[A(\delta M_z)^2 + B(\delta M_z)
|\vec{M}_{\perp}|^2 + C|\vec{\nabla}\vec{M}_{\perp}|^2\right.\nonumber\\
&~&\left.+ D|\vec{M}_{\perp}|^2 + u_{\perp}|\vec{M}_{\perp}|^4\right]
\label{FreeTransition}
\end{eqnarray}
where we have defined $A \equiv g_z + 6M_0^2u_z$, $B \equiv 2M_0u_{\perp z}$, and $ D \equiv t_{\perp}+u_{\perp z}M_0^2$.
Initially, (i.e., when $H$ is large and negative), we have $A, D>0$.

Now let us consider increasing the value of $H$ gradually while fixing all other parameters (e.g., the temperature).
Increasing $H$ decreases $M_0$, hence decreasing the coefficient $D$. We find that for $t_{\perp}<0$, when $H$ reaches \begin{eqnarray} H_c = \left[-g_z+{2u_z\over u_{\perp z}}t_{\perp}\right] \sqrt{-t_{\perp}\over u_{\perp z}} , \label{AKDF} \end{eqnarray} $D=0$.
$H=H_c$ defines the tilting locus $AKDF$ in the figure. The locus $DE$ is the mirror image of $DF$ about the $t_{\perp}$ axis.
For $H>H_c$, $D<0$,
and the magnetization tilts away from $-\hat{z}$ such that its transverse component $\vec{M}_{\perp}$ becomes nonzero. For small tilting, the tilt angle $\theta\propto|\vec{M}_{\perp}|/M_0$. Therefore, $\vec{M}_{\perp}$ is the order parameter of this transition, and $\theta$ is proportional to its magnitude.

However, if the coefficient $A$ changes sign before $D$ does, the untilted phase becomes unstable. This happens at $H=H_i^u\equiv-{g_z\over 3}\sqrt{-2g_z\over 3u_z}$. Thus for the tilting transition to occur we must have $H_c<H_i^u$.
In terms
of the parameters in model (\ref{H with field}), this condition can be written as $t_{\perp}<t_{max}\equiv{u_{\perp z}\over {6u_z}}g_z$.
For $t_{\perp}>t_{max}$, increasing $H$ only leads to the complete flipping of the magnetization from $-\hat{z}$ to $\hat{z}$ without any tilting.
This flipping occurs on the locus $AJ$ in the figure.
Assuming $t_{\perp}<t_{max}$, near the tilting transition we can eliminate the degree of freedom $\delta M_z$ from the free energy Eq. (\ref{FreeTransition}).
This gives:
\begin{eqnarray}
F &=& {1\over 2}\int d^dr~ \left[C|\vec{\nabla}\vec{M}_{\perp}|^2+
D|\vec{M}_{\perp}|^2 \right.\nonumber\\
&~&\left.+ \left(u_{\perp}-{B^2\over 4A}\right)
|\vec{M}_{\perp}|^4\right]\, ,
\label{Simple FreeTransition}
\end{eqnarray}
where $D\sim H_c-H$ near the transition.
For the tilting transition to be
continuous, at $H=H_c$ the coefficient of the quartic term has to be positive, which leads to
\begin{eqnarray}
t_{\perp}<{g_zu_{\perp}u_{\perp z}\over 6u_{\perp}u_z-u_{\perp z}^2}.
\end{eqnarray}
This condition can be satisfied in the region on the left side of the point $K$ in the figure. Thus we conclude that a continuous tilting transition is possible on the locus $KDF$. Since $\vec{M}_{\perp}$ is a two-component vector, and since the model Eq. (\ref{H with field}) is invariant under rotations of ${\vec M_{\perp}}$ in its X$Y$ plane, the universality class of the tilting transition is that of the three-dimensional $XY$ model.
Recognizing this leads to the critical exponents for this transition quoted in Eqs. (\ref{tilt_angle}), (\ref{susc_corr_length}), and (\ref{Weak}) at the beginning of this paper.

Further increasing $H$ after the continuous tilting transition eventually leads to an instability of the tilted phase.
The determinant of the Hessian matrix of the Landau free energy Eq. (\ref{H with field}) vanishes at this instability limit, which, after some algebra, we find is at \begin{eqnarray} H={1\over u_{\perp}\sqrt{4u_{\perp}u_z-u_{\perp z}^2}} \left({t_{\perp}u_{\perp z}-2g_zu_{\perp}}\over 3\right)^{3\over 2}.
\label{instability}
\end{eqnarray}
This instability limit is illustrated by the locus $KBC$ in the figure.

Our main results
are summarized by the phase diagram in
$H$-$t_{\perp}$ plane illustrated in the figure. The other features of this phase diagram, including the locii of all the phase boundaries and special points, can be obtained by straightforward minimization of the Landau free energy. Details will be given in a future publication\cite{future}.

Since the lifetimes of the metastable states
are finite,  the transitions between them will be observable only if the experiment is conducted within their lifetimes.

Because the Landau coefficients $g_z$, $t_{\perp}$, $u_z$, $u_{\perp z}$ are all temperature dependent, the tilting transition can be induced by tuning the temperature as well.
For convenience, in what follows we consider the tilting transition induced by fixing $H$ and varying $T$ only. The corresponding results for the opposite case (i.e., fixing $T$ and varying $H$) can be derived in essentially the same way.

Fluctuations of $\delta M_z$ are greatly affected near the tilting transition by the critical fluctuations of $\vec{M}_{\perp}$.
To see this, note that the model Eq. (\ref{FreeTransition}) implies that the truly massive quantity is the combination $\delta M_z+
B|\vec{M}_{\perp}|^2/2A$. Therefore,
\begin{eqnarray}
\left<\delta M_z\right>=-{B\over 2A}
\left<|\vec{M}_{\perp}|^2\right> =
A_{L\pm}|T-T_c|^{1-\alpha}+\mbox{constant}\nonumber\\
\label{alpha0}
\end{eqnarray}
where the $\pm$ subscript distinguishes coefficients above and below $T_c$, respectively.
These coefficients are different, and non-universal; however, their {\it ratio} $A_{L+}\over A_{L-}$ {\it is} universal, and identical to the analogous ratio of the specific heat coefficients above and below $T_c$.

In deriving the above equation
we have used the well-known result \cite{alpha, berg} from the theory of critical phenomena:
\begin{eqnarray}
\left<|\vec{M}_{\perp}|^2\right> =
A_{T\pm}|T-T_c|^{1-\alpha}+\mbox{constant}\, , \label{alpha} \end{eqnarray} and defined $A_{L\pm}=-BA_{T\pm}/2A$.

Now we can calculate the average value of $M_z$ using $\left<M_z\right>=-M_0+\left<\delta M_z\right>$. This gives
\begin{eqnarray} \left<M_z\right>
=-M_0(H,T)+A_{L\pm}|T-T_c|^{1-\alpha}+\mbox{constant}.
\label{M_z sing}
\end{eqnarray}
Thus we see that $M_z$ exhibits a singularity at the tilting transition, despite the fact that $M_z$ itself is {\it not} the order parameter for this transition. Indeed, this singularity (\ref{M_z
sing}) of $M_z$ is very similar to, and arises from the same mechanism as, the well-known\cite{alpha} singularity of the lattice constant in a compressible ferromagnet at the ferromagnet-paramagnet transition\cite{berg}.

Right at the critical point, where  $D=0$, the model Eq. (\ref{FreeTransition}) becomes very similar to that for an isotropic ferromagnet ordered with $\vec M \parallel {\hat z}$. In such an isotropic ferromagnet, the longitudinal susceptibility $\chi_z$ is divergently renormalized by the transverse fluctuations\cite{Mazenko}.
We expect a similar anomaly here, with the crucial difference that, in the isotropic case, rotational invariance requires that the coefficients in model Eq. (\ref{FreeTransition}) satisfy $A=B=4u_{\perp}$, while in our problem, $B$ and $u_{\perp}$ are independent, since rotational invariance is broken even when $D=0$.
Similar issues arise in the smectic-$A$-smectic-$C$ phase transition in anisotropic environments \cite{Pelcovits, CT}.

We calculate this anomaly as follows: In real space $\chi_z(\vec{r}, \vec{r}\,')$ gives the change in $\langle M_z(\vec{r})\rangle$ in response to an external field which is along $\hat{z}$ and
at $\vec{r}\,'$. That is,
\begin{eqnarray}
T\chi_z(\vec{r}, \vec{r}\,')
&=&\langle [\delta M_z(\vec{r})-\langle \delta M_z(\vec{r})\rangle][\delta M_z(\vec{r}\,')-\nonumber\\
&&\langle \delta M_z(\vec{r}\,')\rangle]\rangle.
\end{eqnarray}
Noting that the combination $\delta M_z+{B|\vec{M}_{\perp}|^2\over 2A}$ is massive, we write $\delta M_z=-{B|\vec{M}_{\perp}|^2\over 2A}$; then,
in Fourier space \begin{eqnarray} \chi_z(\vec{q})={B^2\over 4A^2T}G(\vec{q}), \label{relation3} \end{eqnarray} where $G(\vec{q})$ is the Fourier
transform of the correlation function $\left<|\vec{M}_{\perp}(\vec{r})|^2
|\vec{M}_{\perp}(\vec{0})|^2\right>-\left<|\vec{M}_{\perp}|^2\right>^2$.
Standard scaling arguments\cite{lubensky_book} imply that near the transition
\begin{eqnarray}
G(\vec{q})\approx\mbox{constant}+\left\{
\begin{array}{ll}
C_{\pm}'q^{\kappa}, &q\gg\xi^{-1}\\
C_{\pm}'\xi^{-\kappa}, &q\ll \xi^{-1}
\end{array},
\label{Gq}
\right.
\end{eqnarray}
where $\kappa=-\alpha/\nu$, and $ C_{\pm}'=A_{T\pm}$.
Plugging (\ref{Gq}) into (\ref{relation3}) leads to the result (\ref{Weak}).

The results described earlier for the tilting transition in cubic and orthorhombic ferromagnets can be derived by
applying the type of analysis used above for hexagonal ferromagnets to Landau theories that respect the (different)
symmetries for those cases. This will be discussed in a future publication\cite{future}.

In summary, we have studied the field induced phase transitions between ordering states in anisotropic hexagonal ferromagnets.
Some of these transitions are signaled by a tilting of the orientation of the magnetization; in some cases these tilting
transitions occur between {\it metastable} states. We also found the universality classes of these transitions. Similar
scenarios can occur for a variety of crystal symmetries and field orientations\cite{future}.

The ideas presented here have implications for  ferromagnetic superconductors\cite{FluxE, TR} which exhibit a spontaneous-magnetization-induced flux lattice.
Since these materials are anisotropic ferromagnets, they should  also undergo a tilting transition in an applied external field.
The coupling between the magnetization and the flux lines will change the tilting  transition.
We leave this interesting problem for future work\cite{future}.

\end{document}